\documentclass[10pt, onecolumn]{IEEEtran}
\usepackage{makeidx}
\usepackage{amsmath}
\usepackage{epsfig}
\usepackage{amssymb}
\usepackage{textcomp}
\usepackage{pstricks,pst-text,psfrag}
\usepackage{algorithm}
\usepackage{algorithmic}
\usepackage{datetime}
\usepackage{setspace}
\usepackage{watermark}
\usepackage{graphicx}

\usepackage[caption=false]{subfig}

\usepackage{cases}
\usepackage[noadjust]{cite}

\newcommand{\mbf}{\mathbf}

\newcommand{\bzero}{\mbf{0}}

\def\b1{{\mathbf 1}}

\def\bepsilon{{\mbox{\boldmath{$\epsilon$}}}}
\def\bgamma{{\mbox{\boldmath{$\gamma$}}}}
\def\bGamma{{\mbox{\boldmath{$\Gamma$}}}}

\def\blambda{{\mbox{\boldmath{$\lambda$}}}}

\def\bSigma{{\mbox{\boldmath{$\Sigma$}}}}

\newcommand{\cN}{\ensuremath{\mathcal{N}}}

\newcommand{\argmin}{\mbox{\rm arg}\min}

\newcommand{\diag}{\mbox{\rm diag}}
\newcommand{\bdiag}{\mbox{\rm bdiag}}

\def\eg{{e.g.,\ }}
\def\ie{{i.e.,\ }}

\newcommand{\betab}{\begin{tabbing}}
\newcommand{\entab}{\end{tabbing}}
\newcommand{\beitem}{\begin{itemize}}
\newcommand{\enitem}{\end{itemize}}
\newcommand{\bea}{\begin{array}}
\newcommand{\ena}{\end{array}}
\newcommand{\beq}{\begin{equation}}
\newcommand{\enq}{\end{equation}}
\newcommand{\beqa}{\begin{eqnarray}}
\newcommand{\enqa}{\end{eqnarray}}
\newcommand{\beqan}{\begin{eqnarray*}}
\newcommand{\enqan}{\end{eqnarray*}}
\newcommand{\beenum}{\begin{enumerate}}
\newcommand{\enenum}{\end{enumerate}}
\newcommand{\DL}{\begin{dashlist}}
\newcommand{\DLE}{\end{dashlist}}

\newcommand{\ba}{{\ensuremath{\mathbf{a}}}}

\newcommand{\bc}{{\ensuremath{\mathbf{c}}}}
\newcommand{\bd}{{\ensuremath{\mathbf{d}}}}

\newcommand{\bh}{{\ensuremath{\mathbf{h}}}}

\newcommand{\bx}{{\ensuremath{\mathbf{x}}}}
\newcommand{\by}{{\ensuremath{\mathbf{y}}}}
\newcommand{\bz}{{\ensuremath{\mathbf{z}}}}

\newcommand{\bA}{{\ensuremath{\mathbf{A}}}}
\newcommand{\bB}{{\ensuremath{\mathbf{B}}}}
\newcommand{\bC}{{\ensuremath{\mathbf{C}}}}

\newcommand{\bF}{{\ensuremath{\mathbf{F}}}}
\newcommand{\bG}{{\ensuremath{\mathbf{G}}}}

\newcommand{\bI}{{\ensuremath{\mathbf{I}}}}

\newcommand{\bP}{{\ensuremath{\mathbf{P}}}}

\newcommand{\bU}{{\ensuremath{\mathbf{U}}}}
\newcommand{\bV}{{\ensuremath{\mathbf{V}}}}
\newcommand{\bW}{{\ensuremath{\mathbf{W}}}}

\def\bSigma{{\mbox{\boldmath{$\Sigma$}}}}

\def\cSigma{\emph{\mbox{\boldmath{$\Sigma$}}}}

\def\btheta{{\mbox{\boldmath{$\theta$}}}}

\def\bnu{{\mbox{\boldmath{$\nu$}}}}

\def\bEta{{\mbox{\boldmath{$\eta$}}}}

\def\wrt{{w.r.t.\ }}
\def\iid{{i.i.d.\ }}
\newcommand{\norm}[1]{\lVert#1\rVert}

\newcommand{\Cramer}{Cram\'{e}r}

\newcounter{remarkCounter}
\stepcounter{remarkCounter}

\newtheorem{theorem}{Theorem}

\def\CO2{{$\text{CO}_2$\ }}

\newcommand{\firstAuthor}       {Raj~Thilak~Rajan}
\newcommand{\secondAuthor}      {Rob-van~Schaijk}
\newcommand{\thirdAuthor}       {Anup~Das}
\newcommand{\fourthAuthor}      {Jac~Romme}
\newcommand{\fifthAuthor}       {Frank~Pasveer}

\newcommand{\theTitle}{Reference-free Calibration in Sensor Networks}
\title{\theTitle}

\author{
\IEEEauthorblockN{
\firstAuthor\IEEEauthorrefmark{1}, 
\secondAuthor\IEEEauthorrefmark{1},
\thirdAuthor\IEEEauthorrefmark{1},
\fourthAuthor\IEEEauthorrefmark{1},
and~\fifthAuthor\IEEEauthorrefmark{1}}\\ 
\IEEEauthorblockA{
\IEEEauthorrefmark{1}
Holst Centre / IMEC-NL, High Tech Campus 31, 5656 AE Eindhoven, The Netherlands \\
}%
\thanks{Corresponding author: R.T.Rajan (e-mail: rajan49@imec.be, rtrajan@ieee.org)}
\thanks{Associate Editor:}%
\thanks{Digital Object Identifier}
}

\begin{document}
\watermark{\footnotesize \textit{\today\} \hspace{10cm}  }}
\watermark{\small{\today\ 
: Submitted to \emph{IEEE} sensor letters}}
\maketitle
\thispagestyle{plain}
\pagestyle{plain}


\begin{abstract} 
Sensor calibration is one of the fundamental challenges in large-scale IoT networks. In this article, we address the challenge of reference-free calibration of a densely deployed sensor network. Conventionally, to calibrate an in-place sensor network (or sensor array), a reference is arbitrarily chosen with or without prior information on sensor performance. However, an arbitrary selection of a reference could prove fatal, if an erroneous sensor is inadvertently chosen. To avert single point of dependence, and to improve estimator performance, we propose unbiased reference-free algorithms. Although, our focus is on reference-free solutions, the proposed framework, allows the incorporation of additional references, if available. We show with the help of simulations that the proposed solutions achieve the derived statistical lower bounds asymptotically. In addition, the proposed algorithms show improvements on real-life datasets, as compared to prevalent algorithms.
\end{abstract}

%

\section{Introduction} Recent advances in technology have enabled the rise of large-scale IoT based networks comprising of numerous sensors, which cater to a diverse portfolio of applications for \eg smart cities, environment monitoring, agriculture, and air quality monitoring \cite{gubbi2013internet}. Sensor calibration is one of the key challenge in such large-scale networks comprising of low-cost and unreliable sensors. Traditional on-field calibration of inaccurate sensors using reference equipment, or lab-based characterization of the sensor (\eg sensor modelling), is a cumbersome and expensive process, particularly for a large number of sensors \cite{spinelle2017b}. Therefore, state of the art calibration techniques employ network-wide calibration (also known as in-place calibration \cite{bychkovskiy2003collaborative},  on-the-fly calibration \cite{hasenfratz2012fly} or macro-calibration \cite{dorffer2017outlier}), to estimate the calibration parameters of the network using on-field measurements. Here, the calibration parameters typically refer to the sensor gains (or sensitivity), sensor offsets (or bias) and/or sensor drift (\ie time-varying offset). In this article, we focus our attention on estimating the gains and offsets of a sensor network.

When a reference is unavailable, typically blind calibration algorithms are enforced (\eg \cite{balzano2007, dorffer2017outlier}). In the blind calibration framework, the sensed physical phenomenon is assumed to lie in a known lower dimensional subspace. This relaxed assumption enables sparsely-deployed sensor networks to calibrate with each other, despite being exposed to different ambient conditions at the same time. However, in the absence of a reference, blind calibration algorithms can only estimate calibration parameters upto to a scalar, and more significantly the offset information is completely lost \cite{balzano2007}. The practical limitation of blind calibration is conventionally overcome by a stronger assumption of homogeneity. In such schemes, the sensor network is considered to be densely deployed (\eg sensor array) and the sensors are implicitly assumed to sample the same homogeneous environment \cite{bychkovskiy2003collaborative,spinelle2017b,hasenfratz2012fly}. Under such ambient conditions, algorithms exploit the temporal correlation between the sensor nodes, and given an arbitrarily chosen sensor within the network, all the sensor gains and offsets can be uniquely estimated. However, the choice of a reference in a network of identical sensors is conventionally arbitrary, and plays a pivotal role in the performance of any algorithm. Moreover, arbitrary selection of references in a network could be fatal, if an erroneous sensor is accidentally chosen. To avert this dependence, we propose reference-free algorithms for a dense sensor network. In contrast to existing methods, we show that the proposed reference-free algorithm avoids single point of failure, and offers more reliability towards estimating the true physical phenomenon. The proposed framework allows the incorporation of single or multiple references (if available), which enables the same algorithm to cater to both reference-free and reference-based scenarios. 


\textit{Overview:} We formulate the problem statement in Section-\ref{sec:preliminaries}, followed by the cost function in Section-\ref{sec:algorithms}. We derive the theoretical lower bounds in Section-\ref{sec:crb}, and proposed algorithms to solve the cost function in Section-\ref{sec:cls-cal}. The choice of reference-free and reference-based solutions are discussed in Section-\ref{sec:constraints}. In Section-\ref{sec:simulations}, we show via Monte-Carlo experiments that the proposed estimators achieve the statistical lower bounds asymptotically. Finally, we validate the performance of these algorithms on an indoor air-quality network comprising of $\text{CO}_2$ sensors.

\textit{Notation:} Scalars are denoted in lowercase and uppercase characters \eg $a,  A$. Vectors and matrices are denoted by bold lowercase characters \eg $\ba$, and in bold uppercase characters \eg $\bA$, respectively. The Kronecker product is indicated by $\otimes$, the transpose operator by ($\cdot)^T$ and $\norm{\cdot}$ is the Euclidean norm. $\b1_N \in \mathbb{R}^{N}$ is a vector of ones, $\bI_N$ is a $N \times N$ identity matrix and $\bzero$ is a matrix of zeros of the appropriate size. $\diag(\ba)$ is a diagonal matrix containing elements of the vector $\ba$ on its diagonal. The matrix $\bdiag(\bA_1, \bA_2, \hdots, \bA_N)$ consists of matrices $\{\bA_i\}^N_{i=1}$ along the diagonal and zeros elsewhere. 
\section{Problem statement} \label{sec:preliminaries} 
 We consider a sensor network of $N$ nodes sensing a unknown phenomenon and producing a real-valued data of length $M$. In the absence of prior information on the sensor behavior, we model the sensor output as a first-order Taylor series. More concretely, the sensor response of the $i$th node is given by $\by_i= \omega_i\bx_i + \phi_i + \bepsilon_i$, where $\bx_i \in \mathbb{R}^{M \times 1}$ is the physical phenomenon sensed by the $i$th node, and $\{\omega_i , \phi_i\}$ are the gain and offset of the corresponding sensor. The stochastic noise plaguing the system is denoted by $\bepsilon_i \sim \cN (0, \sigma^2_i\bI)$, which is assumed to be \iid Gaussian. Rearranging the terms, we have \begin{equation}
    \bx_i= \alpha_i\by_i + \beta_i + \bEta_i 
    = \bV_i\btheta_i + \bEta_i,
\label{eq:inv_linear_model_i}
\end{equation} where $\bV_i = [\by_i, \b1_M] \in \mathbb{R}^{M \times 2}$ is a first order Vandermonde matrix containing the measurements $\by_i$ from the $i$th sensor node, $\btheta_i \triangleq [\alpha_i, \beta_i]^T = [1/ \omega_i, -\phi_i/ \omega_i]^T$ contains the calibration parameter of the $i$th sensor and $\bEta_i = \alpha_i\bepsilon_i \sim \cN(0, \alpha^2_i\sigma^2_i\bI)$ is the noise on the system of equations. Let $\bx=[\bx^T_1, \bx^T_2, \hdots, \bx^T_N]^T \in \mathbb{R}^{NM \times 1}$, then $\forall\ 1 \le i \le N$ we have \begin{equation}
\bx = \bV\btheta + \bEta,
\label{eq:inv_linear_model}
\end{equation} where \begin{subequations}
\begin{align}
    \label{eq:inv_linear_model_V}
    \bV&=       \text{bdiag}(\bV_1, \bV_2, \hdots, \bV_N), \\
    \label{eq:inv_linear_model_theta}
    \btheta&=   [\btheta^T_1,\btheta^T_2, \hdots, \btheta^T_N ]^T, \\
    \label{eq:inv_linear_model_Eta}
    \bEta&=     [\bEta^T_1,\bEta^T_2, \hdots, \bEta^T_N ]^T 
    \sim ~ \cN(\bzero,\bSigma_{\eta}).
\end{align}\label{eq:inv_linear_model_variables}
\end{subequations} Our aim in this article is to estimate the calibration parameters $\btheta$ using the sensor measurement model (\ref{eq:inv_linear_model}).

\subsection{Data model} \label{sec:datamodel} To calibrate the sensors, we assume that the sensors are densely deployed and sampling the same homogeneous environment for a given time duration. More concretely, under noiseless scenario, let $\bar{\bx}_m= [x_{1,m}, x_{2,m},\hdots, x_{N,m}]^T$ be a vector of $N$ measurements from all the nodes at the $m$th instant, then the disagreement between the nodes for the $m$th measurement must be zero \ie $\bP\bar{\bx}_m= \bzero$, where $\bP=N\bI_N - \b1_N\b1^T_N$ is the centering matrix \cite{boydConvexOptimization}. We now extend this definition to all $M$ measurements as \begin{equation}
    (\bP \otimes \bI_M)\bx = \bGamma\bx= \bzero,
    \label{eq:gamma_introduction}
\end{equation} where we define $\bGamma=\bP \otimes \bI_M$. Now, substituting for $\bx$ from (\ref{eq:inv_linear_model}), we have \begin{equation}
    \bGamma\bV\btheta = \bar{\bEta},
    \label{eq:linearized_from_quad_form}
\end{equation} where $\bV$ and $\btheta$ are defined in (\ref{eq:inv_linear_model_variables}), $\bar{\bEta}=\bGamma\bEta \sim\ \cN(\bzero, \bar{\bSigma}_{\eta})$, $\bar{\bSigma}_{\eta}= \bGamma\bSigma_{\eta}\bGamma^T$ and $\bSigma_{\eta}$ is given by (\ref{eq:inv_linear_model_Eta}). Now, let
\begin{equation}
\label{eq:W}
\bW=\big(\bGamma\bSigma^{1/2}_{\eta}\big)^{-1},
\end{equation} be a weighting matrix chosen to pre-whiten the noise s.t. $\mathbb{E}\{(\bW\bar{\bEta})(\bW\bar{\bEta})^T\} \approx \bI$, then we aim to estimate $\btheta$ by solving \begin{equation}
\label{eq:cost_function_0}
    \min_{\btheta}\ 
    0.5\btheta^T\bG\btheta,
\end{equation} where \begin{equation}
    \label{eq:G}
    \bG= \bV^T\bGamma^T\bW^T\bW\bGamma\bV.
\end{equation} In the following section, we propose solutions to solve (\ref{eq:cost_function_0}).

\section{Lower bounds and Algorithms} \label{sec:algorithms} We begin with the observation that the centering matrix $\bP$ spans the Nullspace $\b1_N$ by definition, and subsequently (under noiseless scenario) the matrix product (\ref{eq:G}) is rank deficient by at least $1$. Therefore, the cost function (\ref{eq:cost_function_0}) is ill-posed, and a unique solution does not exist without sufficient constraints on the system \eg reference information in the network. To resolve this problem, we propose a constrained formulation \begin{equation} 
\label{eq:cost_function_1}
\min_{\btheta}\ 0.5 \btheta^T\bG\btheta
\quad \text{s.t.}\ \bC\btheta=\bd,
\end{equation} where $\bC \in \mathbb{R}^{P \times 2N}$ is a constraint matrix comprising of $P$ constraints providing information on references, and $\bd$ is the corresponding response vector. 
\subsection{Constrained \Cramer\ Rao Bounds}\label{sec:crb} We now derive the theoretical lower bound on the variance of an unbiased estimator for (\ref{eq:cost_function_1}). The Fisher information matrix (FIM) of the 2N-variate Normal distribution $\cN(\bGamma\bV\btheta, \bar{\bSigma}_{\eta})$ in (\ref{eq:linearized_from_quad_form}) is $\bF = \bV^T\bGamma^T\bar{\bSigma}^{\dagger}_{\eta}\bGamma\bV$, where $\bV$ is given by (\ref{eq:inv_linear_model_variables}), $\bGamma$ is from (\ref{eq:gamma_introduction}) and $\bar{\bSigma}_{\eta}$ is the noise covariance of $\bar{\bEta}$. Note that $\bar{\bSigma}_{\eta}$ is semi-definite and is rank-deficient, therefore we employ a Moore-Penrose pseudoinverse denoted by ($\dagger$). Now, let $h(\btheta)=\bC\btheta-\bd$ be a nonempty set of constraints \ie consistent, then the Constrained \Cramer\ Rao bound (CRB) on the error variance for an unbiased estimator is given by \begin{equation}
{\mathbb{E}} \left \{ (\hat{\btheta}-\btheta)(\hat{\btheta}-\btheta)^T \right \} 
\ge \cSigma_{\theta}
= \bU(\bU^T\bF\bU)^{-1}\bU^T, 
\label{eq:CCRB}
\end{equation} where $\cSigma_{\theta}$ is the CRB on $\btheta$, $\bU$ is an orthonormal basis for the null space of the gradient of $h(\btheta)$ \cite[Theorem 1]{stoica1998}. In the absence of any constraint, the lowest achievable bound is given by the pseudoinverse of the FIM, \ie
\begin{equation}
    \bSigma_{\theta}= \bF^{\dagger},
    \label{eq:CCRB_unconstrained}
\end{equation} which is the unconstrained CRB \cite{stoica1998,rajanJ1}.

\subsection{Constrained Least squares} \label{sec:cls-cal} We propose a constrained Least squares based estimator, which is a closed-form centralized algorithm to solve the equality constrained cost function (\ref{eq:cost_function_1}).
\begin{theorem}[CLS-CAL, WCLS-CAL] Let $\bB \in \mathbb{R}^{(2N+P) \times (2N+P)}$ be a non-singular matrix of the form \begin{equation}
\bB=
\begin{bmatrix}
\bG 	& \bC^T \\
\bC      			& \bzero
\label{eq:B_kkt}
\end{bmatrix},
\end{equation} then a closed form solution to (\ref{eq:cost_function_1}) is  \begin{equation} \label{eq:KKT}
\hat{\bnu}= \argmin_{\bnu} \norm{\bB\bnu - \bh}^2
= \bB^{-1}\bh,
\end{equation} where $\bG$ is defined in (\ref{eq:G}), $\bh=[\bzero^T, \bd^T]^T$, and $\hat{\bnu}$ is an estimate of $\bnu= [\btheta^T, \blambda^T]^T$, which contains the unknown calibration parameters $\btheta$ and the corresponding Lagrange vector $\blambda \in \mathbb{R}^{P \times 1}$.  
\end{theorem}
\begin{IEEEproof}
See \cite[Section 10.1.1]{boydConvexOptimization}
\end{IEEEproof}  
A solution to (\ref{eq:KKT}) exists for $M\ge2$, and if certain regularity conditions are met (See \cite[Section 10.1.2]{boydConvexOptimization}). If the weighting matrix  in (\ref{eq:G}) is $\bW=\bI$, then (\ref{eq:KKT}) yields the constrained Least squares based calibration (CLS-CAL). Alternatively, with the proposed weighting matrix (\ref{eq:W}),  (\ref{eq:KKT}) is the weighted constrained Least squares based calibration (WCLS-CAL). It is worth noting that the WCLS-CAL estimate is the minimum variance unbiased estimate \cite{kay1993}, which achieves the derived theoretical lower bound asymptotically (\ref{eq:CCRB}).

\subsection{Choice of constraints}
\label{sec:constraints} The performance of the estimators for (\ref{eq:cost_function_1}) will rely both on the sensor data, and on the choice of the constraints levied upon the system \ie choice of known references in the network. Although our focus in this article is on reference-free solutions, it is worth noting that the constraint matrix $\bC$ can be constructed to cater to both reference-based and reference-free scenarios.
\subsubsection{Reference-based calibration} A naive solution to (\ref{eq:cost_function_1}) is to arbitrary assume one of the nodes as a reference node, which yields the following constraint matrix \begin{equation} 
\label{eq:oneConstraint}
\bC= \bc^T_i \otimes \bI_2,\, \bd= [\alpha_i, \beta_i]^T, \end{equation} where $\bc_i \in \mathbb{R}^{N\times 1}$ is a vector of $0$s, with $1$ on the $i$th node indicating the reference. This is implicitly employed when solving conventional reference based calibration algorithms \cite{spinelle2017b,hasenfratz2012fly}. Along similar lines, the constrained formulation can be extended to serve multiple references if available.


\subsubsection{Reference-free calibration} In pursuit of a data independent constraint, we propose the sum constraint \begin{equation}
\label{eq:sumConstraint}
\bC = \b1^T_N \otimes \bI_2,\, \bd= [N, 0]^T.
\end{equation} In words, the sum constraint proposes a virtual reference, whose calibration parameters are a mean of all the calibration parameters of the sensors in the network. For an ideal sensor $\{\alpha, \beta\} = \{1,0\}$ by definition, and hence the choice of the response vector $\bd$. The sum constraint is particularly suited for large networks, in which the number of good quality sensors outweigh the number of substandard sensors. The optimality of this constraint cannot be concretely proven for all scenarios, however we show via Monte Carlo simulations and experiments on real dataset, that the sum constraint achieves near optimal performance. Furthermore, this constraint has been successfully employed in other applications, for \eg clock synchronization \cite{rajanJ1} and gain calibration in radio astronomy \cite{wijnholdsConstrained06}.

\section{Simulations} \label{sec:simulations} We now investigate the performance of the proposed solutions on both synthetic and real data sets. We apply the Least squares algorithm (\ref{eq:KKT}) using constraint (\ref{eq:oneConstraint}) to obtain a single reference-based solution, and employ the constraint (\ref{eq:sumConstraint}) for the reference-free solution. We do not evaluate the performance of multiple-references due to space limitations.
We generate a synthetic data set consisting of $N=10$ sensors, where each sensor measures $M$ samples of a source which is linearly varying from $10$ to $1000$ units. We assume that the gain and offset of the sensors are Gaussian distributed around the mean values of $1$ and $0$ respectively. In addition to the sensor discrepancies, the sensor data is corrupted with i.i.d Gaussian noise, \ie $\bepsilon_i \sim\ \cN(0, \sigma_i^2\bI)$ where the variance is arbitrarily chosen in the range of $[0, 20]$. We use the  root mean square error (RMSE) metric as a performance criterion, defined as RMSE($\btheta,\hat{\btheta}$)=  $\sqrt{N^{-1}_{exp}\bSigma^{N_{exp}}_{n=1} \norm{\hat{\btheta}(n)-\btheta}^2}$, where $N_{exp}$ is the total number of Monte-Carlo runs and $\hat{\btheta}(n)$ is an estimate of the unknown $\btheta \in \mathbb{R}^{2N \times 1}$ from the $n$th experiment. The RMSEs are plotted against the averaged root \Cramer\ Rao bound (RCRB), which is the square root of $\text{trace}(\bSigma_{\theta})$ where $\bSigma_{\theta}$ is from (\ref{eq:CCRB}) or (\ref{eq:CCRB_unconstrained}). 
\begin{figure}[ht]
\centering
{   
    \psfrag{tL}[cb]{\small }
	\psfrag{xL}[cb]{\small Number of measurements ($M$)}
	\psfrag{yL}[cb]{\small RMSE($\btheta,\hat{\btheta}$)}
	\psfrag{la}[lb]{\scriptsize Ref.-based CLS-CAL }
	\psfrag{lb}[lb]{\scriptsize Ref.-based WCLS-CAL}
	\psfrag{lc}[lb]{\scriptsize Ref.-based RCRB}
	\psfrag{ld}[lb]{\scriptsize Ref.-free CLS-CAL }
	\psfrag{le}[lb]{\scriptsize Ref.-free WCLS-CAL}
	\psfrag{lf}[lb]{\scriptsize Ref.-free RCRB}
	\psfrag{lg}[lb]{\scriptsize Unconstrained RCRB}
    \subfloat{\label{subfig-1:dummy}%
    \includegraphics[width=0.41\textwidth]{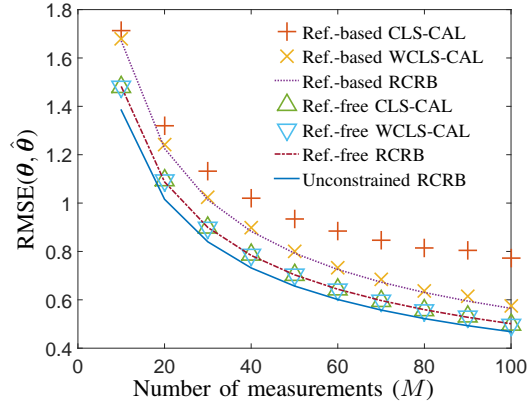}}
} 
\caption{Monte-Carlo simulation: The RMSE error of the proposed estimators are plotted alongside the derived RCRBs for both reference-free and reference-based scenarios, against varying number of measurements $M$ at each sensor node.}
\label{fig:monteCarlo}
\end{figure}

\figurename{\ref{fig:monteCarlo}} shows the RMSEs of the proposed estimators for varying number of sensor measurements $M$, over $N_{exp}=1000$ Monte-carlo runs. The reference-based algorithms (\ref{eq:KKT}) are simulated using (\ref{eq:oneConstraint}), and the respective RMSEs are plotted along with the single reference constrained RCRB (\ref{eq:CCRB}). In contrast to the classical Least squares solution, the WCLS-CAL achieves the RCRB asymptotically as expected. We also simulate the constrained algorithms for reference-free scenarios, \ie with the sum-constraint (\ref{eq:sumConstraint}), which have a near identical performance, and show an improvement in comparison to the single-reference based solutions. The proposed reference-free estimators achieve the theoretical lower bounds asymptotically, and the corresponding CRB is almost comparable to the unconstrained RCRB (\ref{eq:CCRB_unconstrained}). 


\begin{figure*}[!ht]
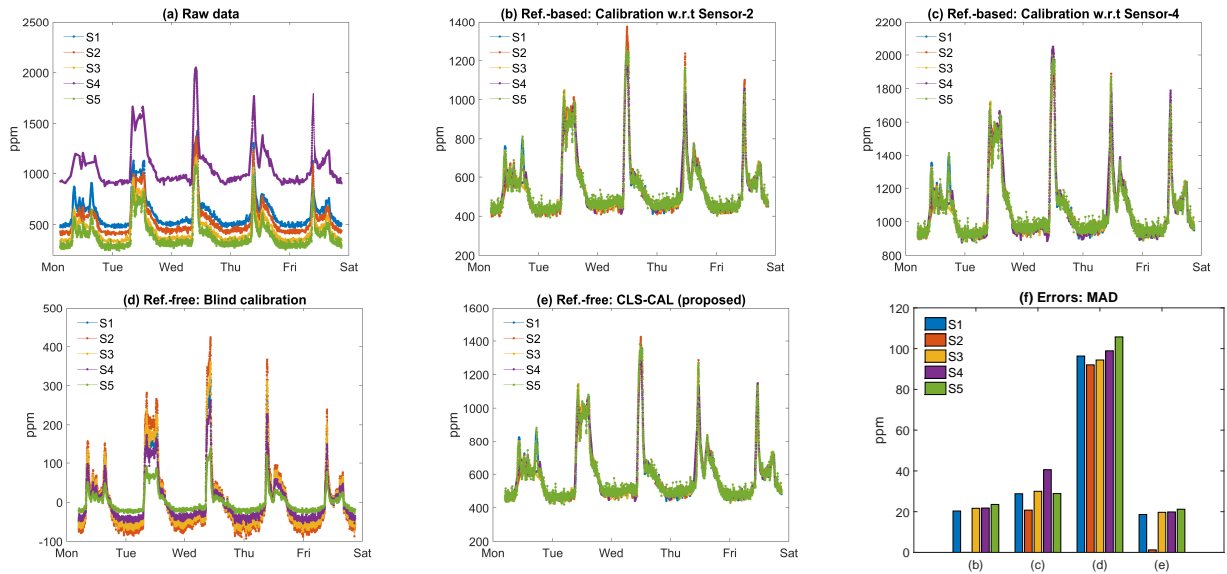

\centering
{   
    \subfloat{
    \includegraphics[width=0.28\textwidth]{data_raw.eps}}
}
\;
{
    \subfloat{
    \includegraphics[width=0.28\textwidth]{data_good_ref.eps}}
}
\;
{
    \subfloat{
    \includegraphics[width=0.28\textwidth]{data_bad_ref.eps}}
}
{
    \subfloat{
    \includegraphics[width=0.28\textwidth]{data_balzano.eps}}
}
\;
{
    \subfloat{
    \includegraphics[width=0.28\textwidth]{data_central_sum.eps}}
}
\;
{
    \subfloat{
    \includegraphics[width=0.26\textwidth]{bar_mad.eps}}
} \caption{Indoor $\text{CO}_2$ sensor network: (a) shows the data from a network of $5$ co-located \CO2 sensors deployed in an office room, where the sensors sample a homogeneous environment. S2 is calibrated and indicates the `ideal' reference.  The reference-based calibrated datasets are show in (b) and (c), where S2 and S4 are used as a reference respectively. Reference-free results are shown in (d) and (e), which are the output of \cite{balzano2007} and the proposed reference-free CLS-CAL (\ref{eq:KKT}) respectively. The MAD errors of (b)-(e) w.r.t S2 are shown in (f)} 
\label{fig:dataset_algo}
\end{figure*}

\begin{table}
\centering
\caption{MAEs of  \figurename{\ref{fig:dataset_algo}(b)-(e)} \wrt reference S2}
\scriptsize{
\begin{tabular}{l|l|l|l|l|l|}
\cline{2-6}
                          & S1 & S2 & S3 & S4 & S5 \\ \hline
\multicolumn{1}{|l|}{\figurename{\ref{fig:dataset_algo}}(b)} & 20.33    & 0.00     & 21.66    & 21.79    & 23.56    \\ \hline
\multicolumn{1}{|l|}{\figurename{\ref{fig:dataset_algo}}(c)} & 526.87   & 526.87   & 526.87   & 526.87   & 526.87   \\ \hline
\multicolumn{1}{|l|}{\figurename{\ref{fig:dataset_algo}}(d)} & 544.49   & 544.49   & 544.49   & 544.49   & 544.49   \\ \hline
\multicolumn{1}{|l|}{\figurename{\ref{fig:dataset_algo}}(e)} & 43.02    & 41.83    & 45.89    & 44.00    & 43.55    \\ \hline
\end{tabular}}
\end{table}

We now focus our attention on a real-world dataset, obtained from $5$ NDIR (Non-dispersive infrared) \CO2 sensors which are co-located in an office environment during a full working week. The time-series obtained from these sensors \{S1, S2, S3, S4, S5\} are shown in  \figurename{\ref{fig:dataset_algo}(a)}, where the diurnal activity of the environment is evident. During the day, the sensors indicate room occupancy, and in the absence of employees in the night, the sensors sample the atmospheric \CO2, which is approximately $410$ppm (see https://www.co2.earth/). Our aim is to investigate the final result of the proposed algorithms on the $\text{CO}_2$ calibrated dataset. To this end, we estimate the calibration parameters of individual sensors, and apply it back on the dataset using (\ref{eq:inv_linear_model_i}) to obtain a calibrated dataset. 

In our dataset, S2 is lab-calibrated which we use as an `ideal' reference for validation, whereas S4 is the most inaccurate. The effect of choosing a healthy reference (S2) or an erroneous reference (S4) is shown in \figurename{\ref{fig:dataset_algo}(b)} and \figurename{\ref{fig:dataset_algo}(c)} respectively, where reference-based algorithm using CLS-CAL has been applied. The sensor offset of the calibrated sensors in \figurename{\ref{fig:dataset_algo}(b)} is around the expected range of $410$ppm, however in case S4 is chosen inadvertently, the sensor offsets of all the sensors will be wrongly corrected to $\approx 1000$ppm. In case of reference-free scenarios, we apply the classical blind-calibration algorithm on the raw dataset, where we assume that the underlying physical phenomenon resides in a rank-1 subspace \cite{balzano2007}. The resulting output in \figurename{\ref{fig:dataset_algo}}(d) shows the sensor offsets are centered around $0$, and additionally the gains are estimated only upto a scaling factor, in the absence of a reference. In contrast, our proposed reference-free algorithm based on the sum-constraint (\ref{eq:sumConstraint}) shows comparable results to the reference-based solution shown in \figurename{\ref{fig:dataset_algo}(b)}. 

Now, let $\bz_i$ be the calibrated data vector of length $L$ obtained from a given solution for the $i$th sensor, and let $\bz$ indicate the raw reference data from S2, then the absolute error is given by $\bgamma= |\bz_i-\bz| \in \mathbb{R}^{L \times 1}$. Following immediately, the mean absolute error is $\text{MAE}(\bgamma) = L^{-1}\bgamma$, which for the calibrated datasets in \figurename{\ref{fig:dataset_algo}(b)-(e)} is given in Table-1.  Not surprisingly, the prior knowledge of a healthy reference in S2 yields the lowest MAEs for all the sensors, and $0$ for S2 itself. The MAE of the proposed reference-free solution is a factor $2$ more than the `ideal' S2-based calibration. However, the MAEs of the blind calibration and S4-based calibration are an order magnitude higher, primarily due to their respective sensor offset errors, as clearly seen in \figurename{\ref{fig:dataset_algo}(c)-(d)}. For a fair comparison of the results, we choose to exclude the offset error, by estimating the mean absolute deviation (MAD) of the $\bgamma$ \ie  $\text{MAD}(\bgamma) = \bgamma- L^{-1}\bgamma^T\b1_L$, which is shown in \figurename{\ref{fig:dataset_algo}(f)}. The `ideal' S2-based calibration solution yields the lowest MADs for all the sensors and the choice of S4 marginally increases these errors. However, the errors are significantly large for blind calibration techniques, since the sensor gains are estimated only up to a scalar in the absence of a reference. Finally, the proposed reference-free solution for this dataset shows comparable results \wrt to the `ideal' S2-based solution. The reduced MAEs and MADs for our proposed reference-free solution, is largely due to the fact that the number of healthy sensors clearly outweigh the number of substandard sensors in our dataset \figurename{\ref{fig:dataset_algo}(a)}. Alternatively, if multiple references are available in the network, then the errors are further expected to reduce, and consequentially improve estimator performance \cite{rajanJ1}.  
\section{Conclusions} \label{sec:conclusions} In this article, we presented closed-form algorithms to calibrate a densely populated sensor network in the absence of reference. The proposed framework caters to both reference-based and reference-free scenarios, and hence additional reference(s) can be incorporated if available. Simulation results show that the proposed estimators achieve the statistical lower bounds asymptotically. Experiments conducted on real-life datasets reveal the benefits of using reference-free calibration techniques in densely deployed sensor networks. The proposed solution can be naturally extended to a time-varying state-space model, and subsequently long-term calibration can be achieved using adaptive filters \cite{kay1993}. Furthermore, the proposed constrained Least squares algorithms can be readily distributed to support resource-constrained processing, and to ensure efficient communication \cite{boyd2011distributed}.

\bibliographystyle{IEEEtran}
\bibliography{ref}

\begin{thebibliography}{10}
\providecommand{\url}[1]{#1}
\csname url@samestyle\endcsname
\providecommand{\newblock}{\relax}
\providecommand{\bibinfo}[2]{#2}
\providecommand{\BIBentrySTDinterwordspacing}{\spaceskip=0pt\relax}
\providecommand{\BIBentryALTinterwordstretchfactor}{4}
\providecommand{\BIBentryALTinterwordspacing}{\spaceskip=\fontdimen2\font plus
\BIBentryALTinterwordstretchfactor\fontdimen3\font minus
  \fontdimen4\font\relax}
\providecommand{\BIBforeignlanguage}[2]{{%
\expandafter\ifx\csname l@#1\endcsname\relax
\typeout{** WARNING: IEEEtran.bst: No hyphenation pattern has been}%
\typeout{** loaded for the language `#1'. Using the pattern for}%
\typeout{** the default language instead.}%
\else
\language=\csname l@#1\endcsname
\fi
#2}}
\providecommand{\BIBdecl}{\relax}
\BIBdecl

\bibitem{gubbi2013internet}
J.~Gubbi, R.~Buyya, S.~Marusic, and M.~Palaniswami, ``Internet of things (iot):
  A vision, architectural elements, and future directions,'' \emph{Future
  generation computer systems}, vol.~29, no.~7, pp. 1645--1660, 2013.

\bibitem{spinelle2017b}
L.~Spinelle, M.~Gerboles, M.~G. Villani, M.~Aleixandre, and F.~Bonavitacola,
  ``Field calibration of a cluster of low-cost commercially available sensors
  for air quality monitoring. part b: No, \{CO\} and \{CO2\},'' \emph{Sensors
  and Actuators B: Chemical}, vol. 238, pp. 706 -- 715, 2017.

\bibitem{bychkovskiy2003collaborative}
V.~Bychkovskiy, S.~Megerian, D.~Estrin, and M.~Potkonjak, ``A collaborative
  approach to in-place sensor calibration,'' in \emph{IPSN}, vol.~3.\hskip 1em
  plus 0.5em minus 0.4em\relax Springer, 2003, pp. 301--316.

\bibitem{hasenfratz2012fly}
D.~Hasenfratz, O.~Saukh, and L.~Thiele, ``On-the-fly calibration of low-cost
  gas sensors,'' \emph{Wireless Sensor Networks}, pp. 228--244, 2012.

\bibitem{dorffer2017outlier}
C.~Dorffer, M.~Puigt, G.~Delmaire, and G.~Roussel, ``Outlier-robust calibration
  method for sensor networks,'' in \emph{Electronics, Control, Measurement,
  Signals and their Application to Mechatronics (ECMSM), 2017 IEEE
  International Workshop of}.\hskip 1em plus 0.5em minus 0.4em\relax IEEE,
  2017, pp. 1--6.

\bibitem{balzano2007}
L.~Balzano and R.~Nowak, ``Blind calibration of sensor networks,'' in
  \emph{Proceedings of the 6th international conference on Information
  processing in sensor networks}.\hskip 1em plus 0.5em minus 0.4em\relax ACM,
  2007, pp. 79--88.

\bibitem{boydConvexOptimization}
S.~Boyd and L.~Vandenberghe, \emph{{Convex Optimization}}.\hskip 1em plus 0.5em
  minus 0.4em\relax Cambridge University Press, Mar. 2004.

\bibitem{stoica1998}
P.~Stoica and B.~C. Ng, ``On the {C}ramer-{R}ao {B}ound under parametric
  constraints,'' \emph{{IEEE} Signal Processing Letters}, vol.~5, no.~7, pp.
  177--179, 1998.

\bibitem{rajanJ1}
R.~T. Rajan and A.-J. {van der Veen}, ``Joint ranging and synchronization for
  an anchorless network of mobile nodes,'' \emph{{IEEE} Transactions on Signal
  Processing,}, vol.~63, no.~8, pp. 1925--1940, 4 2015.

\bibitem{kay1993}
S.~M. Kay, \emph{Fundamentals of statistical signal processing: estimation
  theory}.\hskip 1em plus 0.5em minus 0.4em\relax Upper Saddle River, NJ, USA:
  Prentice-Hall, Inc., 1993.

\bibitem{wijnholdsConstrained06}
S.~Wijnholds and A.~J. {van der Veen}, ``Effects of parametric constraints on
  the {CRLB} in gain and phase estimation problems,'' \emph{IEEE Signal
  Processing Letters}, vol.~13, no.~10, pp. 620 --623, 10 2006.

\bibitem{boyd2011distributed}
S.~Boyd, N.~Parikh, E.~Chu, B.~Peleato, J.~Eckstein \emph{et~al.},
  ``Distributed optimization and statistical learning via the alternating
  direction method of multipliers,'' \emph{Foundations and
  Trends{\textregistered} in Machine learning}, vol.~3, no.~1, pp. 1--122,
  2011.

\end{thebibliography}
\end{document}